\documentclass{jetpl}
\usepackage[cp1251]{inputenc}
\usepackage[english]{babel}
\twocolumn

\rus

\DeclareMathOperator{\curl}{curl}

\title{Diffraction radiation from a screen of finite conductivity}

\rtitle{}

\sodtitle{}

\author{D.\,V.\,Karlovets\/\thanks{karlovets@tpu.ru}, A.\,P.\,Potylitsyn}

\rauthor{}

\sodauthor{}

\address{Tomsk polytechnic university, pr. Lenina 30, Tomsk, 634050 Russian Federation}

\dates{}{*}

\abstract{An exact solution has been found for the problem of diffraction radiation appearing when a charged particle moves perpendicularly to a thin finite screen having arbitrary conductivity 
and frequency dispersion. Expressions describing the diffraction and Cherenkov emission mechanisms have been obtained for the spectral-angular forward and backward radiation densities.}

\PACS{41.60.-m}

\begin{document}

\maketitle

As is known, a charged particle uniformly moving in vacuum in the vicinity of optical inhomogeneity emits radiation called diffraction radiation \cite{B-66}. This phenomenon is close in nature to transition
radiation appearing when the particle intersects an infinite boundary between two media. The characteristics of transition radiation at the interface between a
vacuum and a medium with a finite permittivity and frequency dispersion $\varepsilon(\omega)$ were obtained in the pioneering work by Ginzburg and Frank
\cite{G-F}. On the contrary, diffraction radiation was considered primarily for perfectly conducting surfaces (screens, gratings, etc., see \cite{B-00}). 
Diffraction radiation in the X-ray range for the case where the relative permittivity of a screen is close to unity was recently considered in \cite{T,T-1}. 
At the same time, the problem of diffraction radiation generated by a particle moving near the screen (target) with finite conductivity is of both theoretical and applied interest. 
The simplest geometry in this case is the passage of the particle with the dimensionless energy $\gamma = E/mc^2 = 1/\sqrt{1-\beta^2}$ near the $\infty \times a\times b$ 
rectangular screen (see Fig. 1). In this work, we obtain a solution of this problem for the case where the screen thickness $b$ is much smaller than its length $a$. 
For good conductors with $\Imag \varepsilon(\omega) \gg 1$ this constraint is insignificant due to the skin-effect.

Both transition and diffraction radiations belong to the polarization radiation emitted by atoms of a medium under the action of the external field $\bold E^0$ 
of the moving particle. Therefore, the diffraction radiation field is a solution of macroscopic Maxwell’s equations for the vacuum with the polarization current 
on the right-hand side; the current density for the non-magnetic medium has the form:
\begin{equation}
\displaystyle \bold j (\bold r, \omega)_{pol} = \sigma (\bold r, \omega) (\bold E^0 + \bold E^{pol} (\bold {j}_{pol})), \label{1}
\end{equation}
which is an integral equation because the field of polarization radiation is a function of the current density. Here, the conductivity of the medium (screen) is related to the relative permittivity as: 
$\sigma (\bold r, \omega) = (\varepsilon (\bold r, \omega) - 1) \omega / (4 \pi i)$. 
Equation (\ref{1}) can be solved by the iteration method taking into account the smallness of the parameter $\varepsilon -1 \ll 1$, which is valid, for example, for
frequencies above the plasma frequency \cite{T}. This method is inapplicable for lower frequencies and a screen with arbitrary conductivity.

We use another method. With the notation
\begin{equation}
\displaystyle \bold j_{pol}^{(0)} = \sigma (\bold r, \omega) \bold E^0(\bold r, \omega), \label{1a}
\end{equation}
the following equation for the magnetic field of polarization radiation, $\bold H^{pol}$, can be derived from Maxwell’s
equations:
\begin{eqnarray}
\displaystyle \Big (\Delta + \varepsilon (\bold r, \omega) \frac{\omega^2}{c^2}\Big ) \bold H^{pol} (\bold r, \omega ) = - \frac{4 \pi}{c} \Big (\sigma (\bold r, \omega) \curl \bold E^0 
\cr \displaystyle - (\bold E^0 + \bold E^{pol}) \times \nabla \sigma (\bold r, \omega) \Big ). \label{2}
\end{eqnarray}
It can be shown that the solution of Eq.(\ref{2}) for an infinite medium ($\sigma (\bold r, \omega) = \sigma (\omega)$) provides an exact
expression for the Cherenkov radiation field. For the simplest case of inhomogeneity in the form of the infinite plane vacuum–medium interface, the conductivity has the form:
$\sigma (\bold r,\omega) = \Theta (z) \sigma (\omega)$, and it is easy to see that: 
\begin{equation}
\displaystyle (\bold E^0 + \bold E^{pol})\times \nabla \sigma (\bold r, \omega) = \sigma (\omega) \delta(z) (\bold E^0 + \bold E^{pol}) \times \bold n, \label{3}
\end{equation}
where $\bold n = \{0, 0, 1\}$ is the unit vector normal to the interface. Thus, the last term on the right-hand side of (\ref{2}) is nonzero only at the interface, where the
boundary condition of the continuity of the tangential field components, $(\bold E^0 + \bold E^{pol})\times \bold n |_{z = 0} = \bold E^0 \times \bold n$, is satisfied. 
This property is also valid for more complex surfaces (e.g., for the rectangular screen shown in Fig. 1)
and, owing to this property, only the “external” field $\bold E^0$ is retained on the right-hand side of Eq. (\ref{2}). 
Writing the standard representation of the solution of Eq. (\ref{2}), we take into account that the region of integration is
reduced to the region where polarization currents exist (target volume $V_{T}$ in the case of diffraction radiation):
\begin{eqnarray}
\displaystyle \bold H^{pol} (\bold r, \omega ) = \curl \frac{1}{c} \int \limits_{V_{T}} \bold j_{pol}^{(0)} (\bold {r}^{\prime}, \omega) 
\frac{e^{i \sqrt{\varepsilon (\omega )} \omega |\bold r - {\bold r}^{\prime}|/c}}{|\bold r - {\bold r}^{\prime}|} d^3 r^{\prime}. \label{4}
\end{eqnarray}
We emphasize that this expression is an exact solution of Maxwell’s equations and allows us to avoid solving integral equation (\ref{1}).
The second term on the right-hand side of Eq. (\ref{1}) finally results only in a change of the vacuum wavenumber $\omega/c$ to $\sqrt{\varepsilon (\omega)}\omega/c$.
\begin{figure}
\center \includegraphics[width=5.00cm, height=5.50cm]{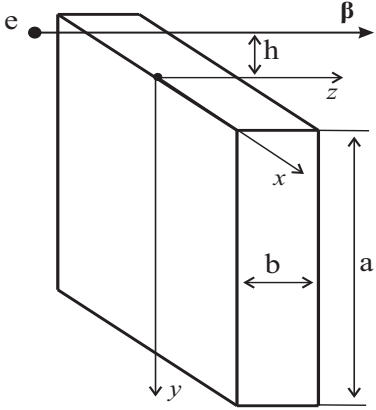}
\caption{\label{Fig1} Fig1. Scheme of generating diffraction radiation.}
\end{figure}

Let us first apply this method to the problem of transition radiation at the infinite interface. In this case, integration is performed over the entire 
half-space $z^{\prime} > 0$, and Eq. (\ref{4}) in the wave zone has the form:
\begin{eqnarray}
\displaystyle && \bold H^{pol} (\bold r, \omega ) = \frac{(2 \pi)^2 i}{c} \frac{e^{i\sqrt{\varepsilon (\omega)}r \omega/c}}{r} \bold k \times \cr 
&& \qquad \qquad \displaystyle \times \int \limits_0^{\infty} dz^{\prime} \bold j_{pol}^{(0)} (\bold k_{\perp}, z^{\prime}, \omega) e^{- i k_z z^{\prime}}. \label{5}
\end{eqnarray}
Here, $\bold k = \omega/c \sqrt{\varepsilon (\omega)}\bold e$, where $\bold e = \bold r/r$, is the wave vector in the medium. The Fourier component of the field of the uniformly moving particle with the charge $e$, which enters into the expression for the current density $\bold j_{pol}^{(0)}$, has the form (below, the frequency dependence of the relative permittivity is implied):
\begin{eqnarray}
\displaystyle && \bold E^{0} (\bold k_{\perp}, z^{\prime}, \omega) = \frac{- i e \gamma}{2 \pi^2 \omega} \frac{e^{i \frac{\omega}{v}z^{\prime}}}{1 + \varepsilon (\beta \gamma e_{\perp})^2}
\{ \sqrt{\varepsilon}\beta\gamma \bold e_{\perp}, \gamma^{-1}\} \label{6}
\end{eqnarray}
The substitution of Eq. (\ref{6}) into Eq. (\ref{5}) provides the expression for the radiation field:
\begin{eqnarray}
\displaystyle && \bold H^{pol} = \frac{e}{\pi c} \frac{\beta}{2}\sqrt{\varepsilon} (\varepsilon -1) \frac{e^{i\sqrt{\varepsilon}r \omega/c}}{r} \{e_y, -e_x, 0\} \cr 
&& \qquad \qquad \qquad \displaystyle \times \frac{1 - \sqrt{\varepsilon } \beta \gamma^2 e_z} {(1 - \beta \sqrt{\varepsilon} e_z) (1 + \varepsilon (\beta \gamma e_{\perp})^2)}, \label{7}
\end{eqnarray}
This formula gives the total field of polarization radiation in the medium. The condition that the term $1 - \beta \sqrt{\varepsilon} e_z$ in the denominator of Eq. (\ref{7}) 
is zero corresponds to the condition of Cherenkov radiation. The components of the vector $\bold e$ in Eq. (\ref{7}) are represented
for backward emission as $\bold e = \{\sin \theta_m \sin \phi, \ \sin \theta_m \cos \phi, - \cos \theta_m\}$ in terms of the polar angle in the medium $\theta_m$.

To determine the radiation field beyond the medium, i.e., in vacuum, it is impossible to directly use the Fresnel refraction laws, because emitting dipoles for good conductors are concentrated near the
interface and the field near the surface does not correspond to the wave zone. In this case, we can use the so-called reciprocity principle \cite{L}:
\begin{eqnarray}
\displaystyle && (\bold E^{pol(vac)}, \bold d^{(vac)}) = (\bold E^{pol(m)}, \bold d^{(m)}), \label{8}
\end{eqnarray}
where $\bold E^{pol(vac)}$ is the desired radiation field in vacuum created by a dipole with the moment $\bold d$ located in the medium and
$\bold E^{pol(m)}$ is the radiation field in the medium created by the same dipole located in vacuum far from the interface. In the problem of transition
radiation, the dipole moment $\bold d$ can be oriented along the only separated direction, which is the $z$ axis. Taking into account that the vector $\bold E^{pol}$ 
is perpendicular to $\bold e$, the radiation field in vacuum appearing in Eq. (\ref{8}) is given by the expression
\begin{eqnarray}
\displaystyle && |\bold E^{pol(vac)}| = \frac{\sin \theta_m}{\sin \theta}|\bold E^{pol(m)}| = \frac{1}{\sqrt{\varepsilon}}|\bold E^{pol(m)}|, \label{9}
\end{eqnarray}
where the known law is used for the relation between the vacuum angle $\theta$ and the angle in the medium $\theta_m$ \cite{L}.
Since the field of the spherical wave in the medium satisfies the equality: $|\bold E^{pol(m)}| = \varepsilon^{-1/2}|\bold H^{pol(m)}|$, 
it remains only to find the magnetic field in the medium for the case where the field of the wave incident on the interface from vacuum is given by Eq.(\ref{7}).
Taking into account the above consideration, from Eq. (\ref{9}), we obtain:
\begin{eqnarray}
\displaystyle && |\bold E^{pol(vac)}|^2 = \frac{1}{|\varepsilon|^2}\Big (|f_H|^2 |H^{pol}_{\perp}|^2 + \cr \displaystyle && \qquad \qquad \qquad \qquad
+ |\sqrt{\varepsilon} f_E|^2 (|H_z^{pol}|^2 + |H_{\parallel}^{pol}|^2)\Big ). \label{10}
\end{eqnarray}
Here,
\begin{eqnarray}
\displaystyle && H^{pol}_{\perp} = H^{pol}_x \cos \phi - H^{pol}_y \sin \phi, \cr
\displaystyle && H^{pol}_{\parallel} = H^{pol}_x \sin \phi + H^{pol}_y \cos \phi \label{11}
\end{eqnarray}
are the components of magnetic field (\ref{7}) perpendicular and parallel to the plane of incidence of the wave on the interface, respectively, and
\begin{eqnarray}
\displaystyle && f_H = \frac{2 \varepsilon \cos \theta}{\varepsilon \cos \theta + \sqrt{\varepsilon - \sin^2 \theta}}, f_E = \frac{2 \cos \theta}{\cos \theta + \sqrt{\varepsilon - \sin^2 \theta}} \label{12}
\end{eqnarray}
are the Fresnel coefficients.
Note that $H^{pol}_{\parallel} = H_z^{pol} = 0$ in the problem of transition radiation. 
In order to determine the radiation intensity in vacuum by means of Eq. (\ref{10}), 
it remains only to express the radiation angles in the medium in terms of the angles in vacuum:
\begin{eqnarray}
\displaystyle && \bold e = \frac{1}{\sqrt{\varepsilon }}\{\sin \theta \sin \phi, \ \sin \theta \cos \phi, - \sqrt{\varepsilon - \sin^2 \theta}\}. \label{13}
\end{eqnarray}
The final expression for the spectral–angular density of the backward transition radiation has the form
\begin{eqnarray}
\displaystyle && \frac{d^2 W}{d\omega d \Omega} = c r^2 |\bold E^{pol(vac)}|^2 = \frac{e^2}{\pi^2 c} \frac{\beta^2 \sin^2 \theta \cos^2 \theta}{(1 - \beta^2 \cos^2 \theta )^2}\cr 
\displaystyle && \qquad \times \Bigg | \frac{(\varepsilon -1) (1 - \beta^2 + \beta \sqrt{\varepsilon - \sin^2 \theta})}{(1 + \beta \sqrt{\varepsilon - \sin^2 \theta}) (\varepsilon \cos \theta
+ \sqrt{\varepsilon - \sin^2 \theta})}\Bigg |^2 \label{14}
\end{eqnarray}
This formula completely coincides with the known Ginzburg–Frank solution obtained by another method (see, e.g., Eq. (116.9) in \cite{L}).
The formula for forward radiation is obtained from this expression by the change $\beta \rightarrow -\beta$.

Let us solve the problem of diffraction radiation. Since the target shown in Fig. 1 is infinite only along the $x$ axis, Eq. (\ref{4}) in the wave zone has the form
\begin{eqnarray}
\displaystyle && \bold H^{pol} (\bold r, \omega ) = \frac{2 \pi i}{c} \frac{e^{i\sqrt{\varepsilon (\omega)}r \omega/c}}{r} \bold k \times \cr 
&& \qquad \displaystyle \int \limits_0^b dz^{\prime} \int \limits_0^a dy^{\prime} \bold j_{pol}^{(0)} (k_x, y^{\prime}, z^{\prime}, \omega) e^{-i k_y y^{\prime} - i k_z z^{\prime}}. \label{15}
\end{eqnarray}
The corresponding Fourier component of the field of the charge is given by the expression:
\begin{eqnarray}
\displaystyle && \bold E^{0} (k_x, y^{\prime}, z^{\prime}, \omega) = \frac{- i e}{2 \pi v} \frac{e^{i \frac{\omega}{v}z^{\prime}}}{\sqrt{1 + \varepsilon (\beta \gamma e_x)^2}}
\{ \sqrt{\varepsilon}\beta\gamma e_x, \cr && \qquad \displaystyle i \sqrt{1 + \varepsilon (\beta \gamma e_x)^2}, \gamma^{-1}\} \ e^{-(y^{\prime} + h) \frac{\omega}{v \gamma}
\sqrt{1 + \varepsilon (\beta \gamma e_x)^2}} \label{16}
\end{eqnarray}
Here, $h$ is the distance between the particle trajectory and the screen. The substitution of Eq. (\ref{16}) into (\ref{15}) yields the expression for the radiation field
\begin{eqnarray}
\displaystyle && \bold H^{pol} = \frac{e \beta \gamma}{4 \pi c}\sqrt{\varepsilon} (\varepsilon -1) \frac{e^{\sqrt{\varepsilon}r \omega/c}}{r} \bold h \frac{e^{i b \frac{\omega}{c}
(\beta^{-1} - \sqrt{\varepsilon}e_z)} - 1}{1 - \beta \sqrt{\varepsilon} e_z }\times \cr \displaystyle && \frac{(e^{-a \frac{\omega}{v \gamma} (i \beta \gamma \sqrt{\varepsilon} e_y 
+ \sqrt{1 + \varepsilon (\beta \gamma e_x)^2})} - 1) e^{-h \frac{\omega}{v \gamma} \sqrt{1 + \varepsilon (\beta \gamma e_x)^2}}}{\sqrt{1 + \varepsilon (\beta \gamma e_x)^2}
(i \beta \gamma \sqrt{\varepsilon} e_y + \sqrt{1 + \varepsilon (\beta \gamma e_x)^2})}, \label{17}
\end{eqnarray}
where
\begin{eqnarray}
\displaystyle && \bold h = \{\gamma^{-1} e_y - i e_z \sqrt{1 + \varepsilon (\beta \gamma e_x)^2}, e_x (\beta \gamma \sqrt{\varepsilon}e_z - \gamma^{-1}), 
\cr \displaystyle && \qquad \qquad \qquad e_x (i \sqrt{1 + \varepsilon (\beta \gamma e_x)^2} - \beta \gamma \sqrt{\varepsilon} e_y)\}. \label{18}
\end{eqnarray}
To determine the radiation field beyond the screen, reciprocity principle (\ref{8}) should also be used. For the reasons given below, we assume that the dipole
moment $\bold d$ is also perpendicular to the interface, i.e., along the $z$ axis in this case. Then, in order to use the Fresnel formulas for the planar interface, 
it is necessary to neglect reflection on the ends of the screen. For this reason, we consider the screen whose thickness $b$ is much smaller than the length $a$, in this case, the 
further consideration is applicable to the angles $\theta$ not too close to $\pi/2$. This condition is insignificant for good conductors, because only a small region of the
medium near the target surface (skin layer) is involved in the formation of radiation.

Let us substitute Eq. (\ref{17}) for the radiation field components into Eq. (\ref{10}) and take into account the transformation of the angles given by Eq. (\ref{13}). 
In addition, taking into account the equality
\begin{eqnarray}
\displaystyle && \Big |\exp\Big \{-a \frac{\omega}{v \gamma} (i \beta \gamma \sin \theta \cos \phi + \cr \displaystyle && \quad + 
\sqrt{1 + (\beta \gamma \sin \theta \sin \phi )^2})\Big \} - 1\Big |^2 = \cr \displaystyle && \qquad = 4 \Big (\sinh^2\Big (\frac{a}{2}\frac{\omega}{v \gamma}
\sqrt{1 + (\beta \gamma \sin \theta \sin \phi )^2}\Big ) + \cr \displaystyle && \qquad + \sin^2 \Big (\frac{a}{2}\frac{\omega}{c} \sin \theta \cos \phi\Big ) \Big )
e^{-a \frac{\omega}{v \gamma} \sqrt{1 + (\beta \gamma \sin \theta \sin \phi )^2}}, \label{19}
\end{eqnarray}
the final expression for the spectral–angular density of backward diffraction radiation after the cancellation of the terms has the form
\begin{eqnarray}
\displaystyle && \frac{d^2 W}{d\omega d \Omega}\Bigg |_{BDR} = c r^2 |\bold E^{pol(vac)}|^2 = \frac{e^2}{\pi^2 c} \beta^2 \cos^2 \theta \Bigg |\frac{\varepsilon -1}{\varepsilon}\times \cr
\displaystyle && \frac{\exp\Big \{i b \frac{\omega}{c}(\beta^{-1} + \sqrt{\varepsilon -\sin^2 \theta})\Big \} - 1}{1 + \beta \sqrt{\varepsilon - \sin^2 \theta}} \Bigg |^2 \times
\cr \displaystyle && \Big (\sinh^2\Big (\frac{a}{2}\frac{\omega}{v \gamma} \sqrt{1 + (\beta \gamma \sin \theta \sin \phi )^2}\Big ) 
+ \cr \displaystyle && + \sin^2 \Big (\frac{a}{2}\frac{\omega}{c}\sin \theta \cos \phi\Big )\Big ) \Big ((1 + (\beta \gamma \sin\theta \sin \phi )^2)\times \cr
\displaystyle && (1 - \beta^2 \cos^2 \theta )\Big )^{-1} \Bigg [ \Big | \frac{\varepsilon}{\varepsilon \cos \theta + \sqrt{\varepsilon -\sin^2 \theta}} \Big (\gamma^{-1} \sin \theta + 
\cr \displaystyle && \sqrt{\varepsilon -\sin^2 \theta} (\beta \gamma \sin \theta \sin^2 \phi + i \cos \phi \times \cr \displaystyle && \sqrt {1 + (\beta \gamma \sin \theta \sin \phi )^2} ) \Big ) 
\Big |^2 + \Big |\frac{\sqrt{\varepsilon}}{\cos \theta + \sqrt{\varepsilon - \sin^2 \theta}}\Big |^2 \times \cr \displaystyle && (\gamma \sin \phi )^2 (1 - \beta^2 \cos^2 \theta ) 
(\sin^2 \theta + |\sqrt{\varepsilon - \sin^2 \theta}|^2)\Bigg ] \cr \displaystyle && \qquad \qquad \qquad \qquad \times e^{-(h + \frac{a}{2}) \frac{2 \omega}{v \gamma} \sqrt{1 + (\beta \gamma \sin \theta \sin \phi)^2}} \label{20}
\end{eqnarray}
In contrast to transition radiation, the formula for forward diffraction radiation cannot be obtained by the simple change $\beta \rightarrow -\beta$, because the change of the
sign of the particle velocity corresponds to the change of the sign of the $z$-component of the Fourier transform of its field (\ref{16}), 
as well as to the change $z^{\prime} \rightarrow -z^{\prime}$. The calculations completely similar to the above calculations give the following expression for forward diffraction radiation:
\begin{eqnarray}
\displaystyle && \frac{d^2 W}{d\omega d \Omega}\Bigg |_{FDR} = \frac{e^2}{\pi^2 c} \beta^2 \cos^2 \theta \Bigg |\frac{\varepsilon -1}{\varepsilon}\times \cr
\displaystyle && \frac{\exp\Big \{i b \frac{\omega}{c}(- \beta^{-1} + \sqrt{\varepsilon -\sin^2 \theta})\Big \} - 1}{1 - \beta \sqrt{\varepsilon - \sin^2 \theta}} \Bigg |^2 \times
\cr \displaystyle && \Big (\sinh^2\Big (\frac{a}{2}\frac{\omega}{v \gamma} \sqrt{1 + (\beta \gamma \sin \theta \sin \phi )^2}\Big ) 
+ \cr \displaystyle && + \sin^2 \Big (\frac{a}{2}\frac{\omega}{c}\sin \theta \cos \phi\Big )\Big ) \Big ((1 + (\beta \gamma \sin\theta \sin \phi )^2)\times \cr
\displaystyle && (1 - \beta^2 \cos^2 \theta )\Big )^{-1} \Bigg [ \Big | \frac{\varepsilon}{\varepsilon \cos \theta + \sqrt{\varepsilon -\sin^2 \theta}} \Big (\gamma^{-1} \sin \theta - 
\cr \displaystyle && \sqrt{\varepsilon -\sin^2 \theta} (\beta \gamma \sin \theta \sin^2 \phi + i \cos \phi \times \cr \displaystyle && \sqrt {1 + (\beta \gamma \sin \theta \sin \phi )^2} ) \Big ) 
\Big |^2 + \Big |\frac{\sqrt{\varepsilon}}{\cos \theta + \sqrt{\varepsilon - \sin^2 \theta}}\Big |^2 \times \cr \displaystyle && (\gamma \sin \phi )^2 (1 - \beta^2 \cos^2 \theta ) 
(\sin^2 \theta + |\sqrt{\varepsilon - \sin^2 \theta}|^2)\Bigg ] \cr \displaystyle && \qquad \qquad \qquad \qquad \times e^{-(h + \frac{a}{2}) \frac{2 \omega}{v \gamma} \sqrt{1 + (\beta \gamma 
\sin \theta \sin \phi)^2}} \label{21}
\end{eqnarray}
This expression includes both diffraction and Cherenkov radiations. The intensity pole under the condition $\beta \sqrt{\varepsilon - \sin^2 \theta} = 1$ corresponds to the latter radiation as
in the case of transition radiation. However, since we consider radiation from the finite-thickness screen, this pole is removable:
\begin{eqnarray}
\displaystyle && \Bigg |\frac{\exp\Big \{i b \frac{\omega}{c}(- \beta^{-1} + \sqrt{\varepsilon -\sin^2 \theta})\Big \} - 1}
{1 - \beta \sqrt{\varepsilon - \sin^2 \theta}} \Bigg |^2  \rightarrow \Big |b \frac{\omega}{\beta c}\Big |^2, \label{21.5}
\end{eqnarray}
in this case, the dependence on $\varepsilon$ disappears. Continuing the comparison with transition radiation, we note that owing to the absence of the term $\sin^2 \theta$ 
in the numerators of Eqs. (\ref{20}), (\ref{21}) the maximum of diffraction radiation is at the angle $\theta = 0$ as expected.
\begin{figure}
\includegraphics[width=8.00cm, height=5.00cm]{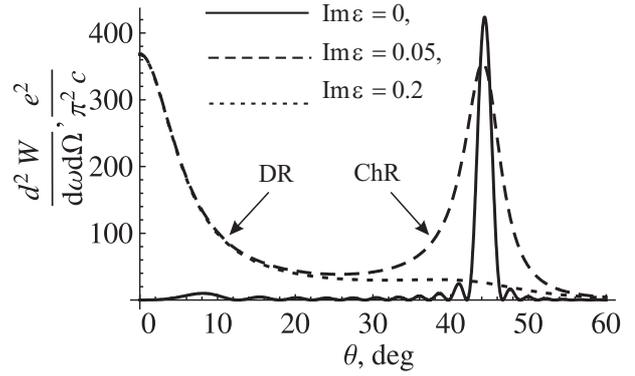}
\caption{\label{Fig2} Fig2. Angular distribution of forward diffraction radiation for various characteristics of the screen substance with the parameters 
$\Real \varepsilon = 1.5, \gamma = 10, a = \infty, b = 50$ mm, $\lambda = h = 1$ mm, and $\phi = 0$. Curves for $\Imag \varepsilon \ne 0$ are magnified by a
factor of 50. A Cherenkov peak, which decreases with increasing $\Imag \varepsilon$, is observed for the angle determined from
the condition $\beta \sqrt{\varepsilon - \sin^2 \theta} = 1$.}
\end{figure}

For the conducting target for which $\Imag \varepsilon(\omega) \gg 1$, the dependence on the screen thickness is absent if $b \gg \lambda/|\sqrt{\varepsilon (\omega)}|$ (skin effect). 
In the limit of perfect conductivity $|\varepsilon| \rightarrow \infty$, Eq. (\ref{20}) is significantly simplified to the form
\begin{eqnarray}
\displaystyle && \frac{d^2 W}{d\omega d \Omega}\Big |_{\epsilon \rightarrow \infty} =  \frac{e^2}{\pi^2 c} \Big (\sinh^2\Big (\frac{a}{2}\frac{\omega}{v \gamma} \sqrt{1 + (\beta \gamma \sin \theta \sin \phi )^2}\Big ) \cr \displaystyle && + \sin^2 \Big (\frac{a}{2}\frac{\omega}{c}\sin \theta \cos \phi\Big )\Big ) \Big ((1 + (\beta \gamma \sin\theta \sin \phi )^2) \times \cr
\displaystyle && (1 - \beta^2 \cos^2 \theta )\Big )^{-1} \Big (1 - \sin^2{\theta}\sin^2{\phi} + \cr
\displaystyle && \qquad + (\beta \gamma \sin \theta \sin \phi )^2 (1 + \cos^2{\theta})\Big ) \times \cr
\displaystyle && \qquad \qquad \qquad \qquad e^{-(h + \frac{a}{2}) \frac{2 \omega}{v \gamma} \sqrt{1 + (\beta \gamma \sin \theta \sin \phi)^2}} \label{22}
\end{eqnarray}

Returning to the problem of the orientation of the emitting dipole in Eq.(\ref{8}), we note that the passage to the limit of a perfect conductor is possible only when
the vector $\bold d$ is perpendicular to the screen surface. If the vector $\bold d$ had the component parallel to the screen
plane, formula (\ref{9}) for the radiation field would contain the term $\propto \cos \theta_m/\cos \theta \sim \sqrt{\varepsilon -\sin^2 \theta}/\sqrt{\varepsilon}$ with the
“excess” degree $\sqrt{\varepsilon}$ in the numerator and the radiation intensity would be infinite in the limit $|\varepsilon| \rightarrow \infty$.
The perpendicular orientation of the dipole to the interface physically means that the thin screen at large distances is a double layer.

The passage to the limit $a \rightarrow \infty$ in Eq.(\ref{22}) gives the following expression for the intensity of backward diffraction radiation from the perfectly conducting
half-plane:
\begin{eqnarray}
\displaystyle && \frac{d^2 W}{d\omega d \Omega} =  \frac{e^2}{4 \pi^2 c} \Big (1 - \sin^2{\theta}\sin^2{\phi} + (\beta \gamma \sin \theta \sin \phi )^2 \times \cr \displaystyle && 
(1 + \cos^2{\theta})\Big ) \Big ((1 + (\beta \gamma \sin\theta \sin \phi )^2) (1 - \beta^2 \cos^2 \theta )\Big )^{-1} \cr
\displaystyle && \qquad \qquad \qquad \qquad e^{- h \frac{2 \omega}{v \gamma} \sqrt{1 + (\beta \gamma \sin \theta \sin \phi)^2}} \label{23}
\end{eqnarray}
This formula completely coincides with that obtained in \cite{J, PLA} by another method and does not coincide with the known solution presented in \cite{K}, 
which is additional evidence that the latter solution is approximate (for more details, see \cite{PLA}). 
\begin{figure}
\includegraphics[width=7.50cm, height=5.00cm]{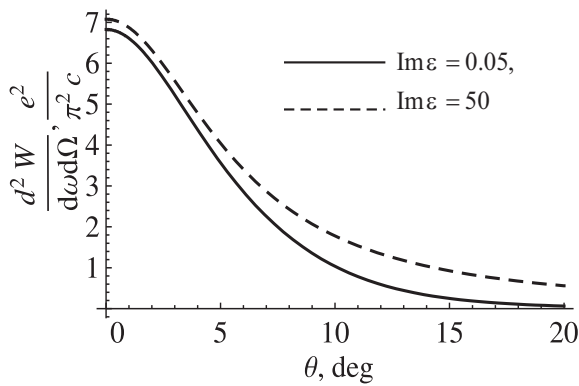}
\caption{\label{Fig3} Fig3. Angular distribution of forward diffraction radiation from the (solid line) absorber and (dashed line) conductor
with the parameters $\Real \varepsilon = 1, \gamma = 10, a = \infty, b = 50$ mm, $\lambda = h = 1$ mm, and $\phi = 0$.}
\end{figure}

Let us present some features of radiation from the target with a finite relative permittivity. For the transparent medium under the Cherenkov condition, 
the intensity of diffraction radiation is low and the main contribution comes from Cherenkov radiation. As the imaginary part of $\varepsilon (\omega)$ increases, the intensity of the
latter radiation decreases rapidly and the angular dependence has the form of a single-peak curve typical
for diffraction radiation (see Fig. 2). It is interesting that the intensity of the forward diffraction radiation
for the screen of an absorbing material with the reflection coefficient close to zero (an absorber with $\varepsilon = 1 + i\,0.05$) almost coincides with a similar dependence for
a substance with the reflection coefficient close to unity (a good conductor with $\varepsilon = 1 + i\,50$), see Fig. 3.
At the same time, the intensity of backward radiation for the absorber is several orders of magnitude lower
than the intensity of forward radiation. The energies emitted in both directions are the same in the limit $\Imag \varepsilon \rightarrow \infty$. Note that the angular dependence 
of diffraction radiation for small $\theta$ angles is independent of $\phi$, whereas Cherenkov radiation is concentrated in the $y0z$ plane perpendicular to the screen plane, see Fig. 4.
\begin{figure}
\includegraphics[width=7.50cm, height=7.00cm]{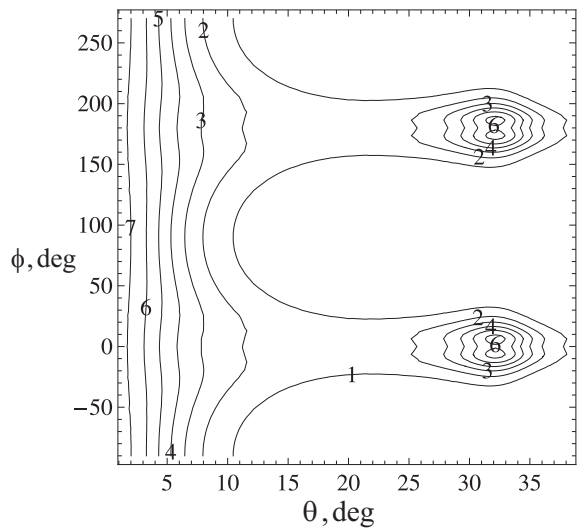}
\caption{\label{Fig4} Fig4. Angular distribution of forward diffraction radiation
taking into account the azimuthal distribution. The parameters are $\Real \varepsilon = 1.3, \Imag \varepsilon = 0.05, \gamma = 10, a = \infty, b = 50$ mm, and $\lambda = h = 1$ mm.}
\end{figure}

To conclude, we again note that the method used to solve the problem is based on the representation of the field of diffraction radiation (transition radiation) as
the radiation field of a polarization current induced in the substance by the field of an external source, which
is a uniformly moving charge in this problem. As shown with the use of boundary conditions, representation (\ref{4}) is also valid for inhomogeneous media with $\varepsilon (\bold r, \omega)$.
Therefore, the problem of the determination of the polarization radiation field in vacuum from the target with a given profile is reduced to the problem of
the refraction of a spherical wave at the screen boundary, for which it is reasonable to use the reciprocity principle.

We are grateful to Drs. A.A. Tishchenko and L.G. Sukhikh for stimulating discussions. This work
was supported in part by the Council of the President of the Russian Federation for Support of Young Scientists and Leading Scientific Schools (project no. NSh4158.2008.2), 
by the Federal Agency for Education of the Russian Federation (project no. 1.81.2006), and by
the Federal Agency for Science and Innovation (project no. 02.740.11.0238).

\vfill\eject

\end{document}